\documentclass[fleqn,usenatbib]{mnras}

\usepackage[T1]{fontenc}

\DeclareRobustCommand{\VAN}[3]{#2}
\let\VANthebibliography\thebibliography
\def\thebibliography{\DeclareRobustCommand{\VAN}[3]{##3}\VANthebibliography}

\usepackage{graphicx}	
\usepackage{amsmath}	
\usepackage{diagbox}

\usepackage[normalem]{ulem}

\newcommand{\TB}{T_{b}}
\newcommand{\TBLIN}{T_{b-\rm lin}}

\newcommand{\MSUN}{{\rm M}_{\odot}}

\newcommand{\TS}{T_{s}}

\newcommand{\TK}{T_{ k}}

\newcommand{\lya}{\rm {Ly{\alpha}}}
\newcommand{\OmegaB}{\Omega_{\rm B}}
\newcommand{\Omegam}{\Omega_{\rm m}}
\usepackage{newtxtext,newtxmath}

\title[Large HI optical depth and 21-cm signal]{
Large HI optical depth and Redshifted  21-cm signal from cosmic dawn \\}

\author[Datta et al.]{
Kanan K. Datta$^{1}$\thanks{E-mail: kanan.physics@presiuniv.ac.in},
Raghunath Ghara$^{2, 3}$, 
Ariful Hoque$^{1}$,
Suman Majumdar$^{4,5}$
\\
$^{1}$Department of Physics, Presidency University, 86/1 College St., Kolkata 700073, India \\
$^{2}$ ARCO (Astrophysics Research Center), Department of Natural Sciences, The Open University of Israel, 1 University Road, PO Box 808, Ra’anana 4353701, Israel\\
$^{3}$ Department of Physics, Technion, Haifa 32000, Israel\\
$^{4}$ Department of Astronomy, Astrophysics and Space Engineering, Indian Institute of Technology Indore, Khandwa Rd., Simrol, MP 453552, India\\
$^{5}$Department of Physics, Blackett Laboratory, Imperial College, London SW7 2AZ, U. K.
}

\date{Accepted XXX. Received YYY; in original form ZZZ}

\pubyear{2015}

\begin{document}
\label{firstpage}
\pagerange{\pageref{firstpage}--\pageref{lastpage}}
\maketitle

\begin{abstract}
The HI 21-cm optical depth ($\tau_b$) can be considerably large as the kinetic and spin temperature of the inter-galactic medium(IGM) is expected to be very low during cosmic dawn.  It will be particularly higher at regions with HI over-density. We revisit the validity of the widely used linearized equation for estimating the HI 21-cm differential brightness temperature ($T_b$) which assumes $\tau_b << 1$ and approximates  $[1-\exp({-\tau_b})]$ as $\tau_b$. We consider two scenarios, one without any additional cooling mechanism or radio background ( referred as  {\it standard} scenario) and  the other (referred as {\it excess-cooling} scenario)  assumes the EDGES like absorption profile and an excess cooling mechanism.  We find that given a measured global absorption signal, consistent with the {\it standard(excess-cooling)} scenario, the linearized equation overestimates the spin temperature by $\sim 5\%(10\%)$. Further, using numerical simulations, we study impact that the large optical depth has on various signal statistics. We observe that the variance, skewness and kurtosis, calculated at simulation resolution ($\sim 0.5 h^{-1} \, {\rm Mpc}$), are over-predicted up to $\sim 30\%$, $30\%$ and $15\%$  respectively for the {\it standard} and up to $\sim 90\%$, $50\%$ and $50\%$ respectively for the  {\it excess-cooling} scenario. Moreover, we find that the probability distribution function of $T_b$ is squeezed and becomes more Gaussian in shape if no approximation is made. The spherically averaged HI power spectrum is over predicted by up to  $\sim 25 \%$ and $80\%$ at all scales for the {\it standard} and {\it excess-cooling} scenarios respectively. 

\end{abstract}




\section{Introduction}
 The redshifted 21-cm signal from atomic neutral hydrogen (HI) in the inter-galactic medium (IGM) is a promising probe for exploring the cosmic dawn when the first luminous sources were formed in the Universe \citep[see][ for reviews]{pritchard12, 2013ASSL..396...45Z}. The signal encodes an enormous amount of information regarding the very first luminous sources, as well as the thermal and ionization state of the intergalactic medium during that epoch \citep{2020MNRAS.493.4728G, ghara2021, 2020MNRAS.498.4178M, 2021MNRAS.501....1G}. 

Currently, a great deal of effort, all over the globe, is underway to detect the signal. Experiments such the EDGES \citep{bowman10}, SARAS \citep{2015ApJ...801..138P, singh18}, LEDA \citep{2012arXiv1201.1700G, leda18}, BigHorns  \citep{2015PASA...32....4S}, SciHi  \citep{2014ApJ...782L...9V}, REACH \citep{8879199} are aiming to detect the sky averaged (global)  HI 21-cm signal at different redshifts, whereas radio interferometric experiments such as the LOFAR \citep{patil17, 2020MNRAS.493.1662M}, MWA \citep{bowman13, barry19, trott20, patwa21}, uGMRT \citep{choudhuri20, pal21, chakraborty21}, HERA \citep{2017PASP..129d5001D}, SKA \citep{2015aska.confE..10M, ghara16} primarily plan to measure the signal statistically. 

The basic measurable quantity in 21-cm cosmology is the HI differential brightness temperature ($\TB$). Experiments mentioned above aim to measure this quantity either directly or statistically using quantities such as the variance, skewness, kurtosis \citep[see e.g.,][]{harker2009, patil2014}, power spectrum \citep[see e.g.,][]{datta12, 2021MNRAS.506.3717R}, bispectrum \citep[see e.g.,][]{majumdar18, 2021MNRAS.502.3800K}. 

The differential brightness temperature depends on the optical depth of HI 21-cm radiation $\tau_b$ as $\TB \propto [1-\exp(-\tau_b)]$ \citep{rybicki86}. The widely used expression for calculating $\TB$ assumes the optical depth to be much smaller than unity and $[1-\exp(-\tau_b)]$ is approximated as $\tau_b$ \citep[see e.g.,][]{park19,  2019MNRAS.487.1101R,  2020MNRAS.492..634G, reis20, 2021JCAP...05..026K}. The optical depth $\tau_b$ inversely varies with the HI spin temperature and is proportional to the HI density \citep{bharadwaj05}. The IGM kinetic and spin temperature are expected to be the lowest just before the heating of the IGM starts during cosmic dawn. Consequently, $\tau_b$ is considerably large immediately before the IGM heating starts, particularly at regions with HI overdensity. The above linear approximation breaks down in such situations. However, we would like to mention that some very recent works do use the exact expression for the differential brightness temperature while modeling the redshifted 21-cm signal from cosmic dawn \citep[e.g.,][]{reis21, xu21}.

In this paper, we investigate the impact of large optical depth on the HI differential brightness temperature and its various statistical quantities such as the variance, skewness, power spectrum during the cosmic dawn. We study this in the `{\it standard}' scenario as well as in a scenario where the IGM is significantly colder in compare to the standard scenario. The second scenario is motivated by the EDGES measurements of global 21-cm signal \citep{EDGES18} and the possibility of cold IGM during cosmic dawn \citep{Barkana18Nature, Munoz18}. Further, we present an analytical formalism that allows us to estimate the HI power spectrum in moderately large HI optical depth regime in a situation when the spin temperature is uniformly distributed. 

Observed global averaged $\TB$ as a function of redshift can be used to estimate the mean spin temperature and IGM temperature during the cosmic dawn \citep{Barkana18Nature}. However, the widely used relation between $\TB$ and spin temperature, which inherently assumes $\tau_b$ to be small, will overestimate the spin temperature. In this work, we study and quantify the impact of a large optical depth on estimating the spin and IGM temperature from observed global 21-cm signal and validity of the widely used linearized equation for  $\TB$.

The outline of the paper is as follows. In section \ref{sec:global}, we present the basic equations for calculating the  HI 21-cm signal. In section \ref{sec:21cm-mean}, we estimate the spin temperature from the EDGES like global 21-cm profile assuming a cold IGM scenario. We present estimated spin temperature for a range of $\TB$ using the linearized and exact equation of $\TB$. Subsequent sections focus on the statistical signal. Section \ref{sec:sim} presents our simulations and various scenarios considered for our study. We present our results in section \ref{sec:res} where different subsections presents the impact of large optical depth on 21-cm images, the probability distribution function of $\TB$, the variance, skewness, kurtosis, and the power spectrum of HI 21-cm brightness fluctuations.  Finally, we summarize our findings in section \ref{sec:con}.

Throughout our analysis, we assume the best-fit values of cosmological parameters consistent with Planck measurements \citep{2014A&A...571A..17P}.

\section{HI 21-cm signal}
\label{sec:global}
The differential brightness temperature corresponding to redshifted HI 21-cm signal at spatial coordinate $\bf{x}$ and redshift $z$ can be written as \citep{rybicki86},
\begin{equation}
    \TB({\bf x}, z)=\frac{\TS({\bf x}, z)-T_{R}(z)}{1+z}\left[1-\exp\{-\tau_{b}({\bf x}, z)\}\right], 
    \label{eq:tb}
\end{equation}
where $T_s$ is the spin temperature corresponding HI hyperfine transition. $T_R(z)$ is the background radiation temperature at 21-cm wavelength at redshift $z$. The HI 21-cm optical depth is denoted as $\tau_{b}$. Note that both $T_s$ and $\tau_{b}$ are functions of spatial coordinate $\bf{x}$ and redshift $z$. The HI 21-cm optical depth can be written as \citep{bharadwaj05},
\begin{eqnarray}
\tau_{b}({\bf x}, z) &=& \frac{4 \, {\rm mK}}{T_s(\mathbf{x}, z)} \frac{\rho_{\rm HI}(\mathbf{x}, z)}{\bar{\rho}_{\rm H}} \left[1-\frac{1+z}{H(z)}\frac{dv}{dr}\right] \nonumber \\ 
&\times& \left( \frac{\Omega_{b0}h^2}{0.02}\right) \left(\frac{0.7}{h}\right)
\frac{H_0}{H(z)}(1+z)^3,
\label{eq:tau}
\end{eqnarray}
where $\rho_{\rm HI}$ is the HI density at $\bf{x}$ and redshift $z$, and $\bar{\rho}_{\rm H}$ is the mean hydrogen density at redshift $z$. The $dv/dr$ is the line of sight component of the rate of change of the peculiar velocity with comoving distance $r$. Other symbols have usual meaning. The global HI 21-cm signal at redshift $z$ can be calculated by averaging $T_{b}(\mathbf{x}, z)$ over all spatial coordinates $\bf{x}$.

 Normally, it is assumed that the HI 21-cm optical depth $\tau_{b} \ll 1$. Keeping up to the linear term of $\exp(-{\tau_b})$ we can write $1-\exp(-{\tau_b}) \approx \tau_b$, and eq. \ref{eq:tb} can be simplified to,
\begin{eqnarray}
T_{\rm b-lin}( \mathbf{x}, z) &=& 4 \, {\rm mK} \left( \frac{\Omega_{b0}h^2}{0.02}\right) \left(\frac{0.7}{h}\right)
\frac{H_0}{H(z)}(1+z)^2  \nonumber  \\ 
&\times& \frac{\rho_{\rm HI}(\mathbf{x}, z)}{\bar{\rho}_{\rm H}(z)} \left[1-\frac{T_R(z)}{T_s(\mathbf{x}, z)} \right]   \left[1-\frac{1+z}{H(z)}\frac{dv}{dr}\right].
\label{eq:tb_lin}
\end{eqnarray}
This equation (or somewhat simplified form) has been widely used both to numerically calculate the HI 21-cm signal and interpret observed signal. 
However, for moderate $\tau_b ({\bf x}, z)$ values eq. \ref{eq:tb} can be better approximated as,
\begin{equation}
   T_{b}({\bf x}, z)=\frac{T_s({\bf x}, z)-T_{R}(z)}{1+z}\left[\tau_b-\tau^2_b/2+0(\tau^3)\right].
    \label{eq:tb-nonlin}
\end{equation}
We see that the differential brightness temperature $T_b$ is a combination of $\tau_b$ and its higher orders. The $\tau^2_b$ and higher orders can be neglected when $\tau_b$ is small enough and, therefore, $T_b$ can be estimated using eq. \ref{eq:tb_lin}. However, before the onset of X-ray heating the IGM could be cold enough to cause $\tau_b$ significantly higher, particularly at over-density regions. In such cases the term $\tau^2_b$ needs to be incorporated. $\tau^2_b \propto \rho^2_{\rm HI}({\bf x}, z)/T^2_s({\bf x}, z)$.  However, incorporating the term $\tau^2_b$ results in reduction of the mean $T_b$ and  fluctuations in the differential brightness temperature. This would impact estimation of statistical quantities such as the variance, skewness, kurtosis, power spectrum of the HI differential brightness temperature. This will be discussed in subsequent sections.

\begin{figure}
	\includegraphics[width=\columnwidth]{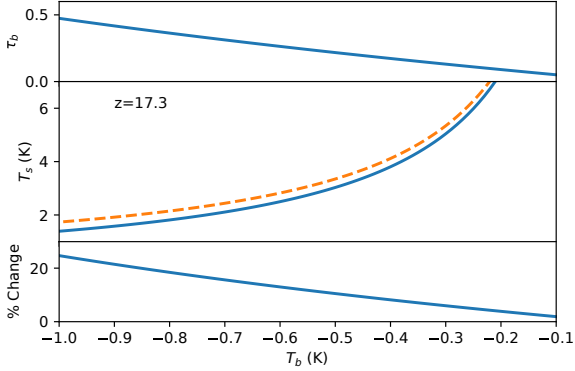}
    \caption{The middle panel shows the estimated spin temperature for a range of measured HI differential brightness temperature $T_b$ at redshift $z=17.3$ using the exact equation \ref{eq:tb} (solid curve) and widely used linear equation \ref{eq:tb_lin} (dashed curve) assuming the mean HI density. The upper panel shows the corresponding HI 21-cm optical depth $\tau_{b}$ estimated using the exact (solid curve) and linear equation (dashed curve). The lower panel shows the percentage change in spin temperature.} 
    \label{fig:tb2ts-mean17}
\end{figure}

\section{21-cm signal at large optical depth and implications of EDGES measurements}
\label{sec:21cm-mean}

We see from eq. \ref{eq:tau} that the HI 21-cm optical depth $\tau_b$ increases when the spin temperature $T_s$ decreases and HI density $\rho_{\rm HI}({\bf x}, z)$ increases. Since $T_s$ remains coupled to the IGM kinetic temperature $T_k$ during the second half of the cosmic dawn and later, the  HI 21-cm differential brightness temperature $T_{b}$ indirectly depends on $T_k$ during that period. In the {\it standard} scenario, $T_k$ goes down as $\sim (1+z)^2$ during the initial phases of cosmic dawn and it is the lowest just before the X-ray heating of the IGM starts. This results in higher HI 21-cm optical depth $\tau_{b}$. For example, $T_k$s are $\sim 3.6$ K, $7$ K at redshifts $z=12$ and $17.2$ respectively as per the {\it standard} scenario if there is no heating \citep{lewis07, Seager2000, datta20}\footnote{https://camb.info/ \\ https://www.astro.ubc.ca/people/scott/recfast.html}. This corresponds to the HI optical depth $\tau_{b} \sim 0.1$ and $\sim 0.08$ at the mean HI density $\bar{\rho}_{\rm HI}$. If the linearized equation (eq. \ref{eq:tb_lin})  is used, it will overestimate the value of $T_{b}$ by $\sim 5\%$ and $\sim 4\%$ at the regions having HI density same the mean HI density. Similarly, the overestimation of $T_{b}$ will rise to $10\%$ and $8\%$ respectively for a typical HI over-density of $\delta_{\rm HI} =1$. Spatial fluctuations in the HI density and kinetic temperature lead to significantly larger effects on the estimation of $T_b$ \citep{pablo20,xu21}.

Conversely, we see that eq. \ref{eq:tb} can be used to accurately estimate the spin temperature $T_s$ for a measured set of differential HI 21-cm brightness temperatures $T_b$ at any spatial coordinate $\textbf{x}$ and redshift $z$. The solid curve in the middle panel of Fig \ref{fig:tb2ts-mean17} shows the estimated spin temperature $T_s$ for a given set of differential brightness temperature $T_b$ assuming mean HI density at redshift $z=17.2$. The choice of $T_b$ range  and redshift is aligned with the EDGES low band measurements \citep{EDGES18}. It also shows (dashed curve) the estimated spin temperature using the linearized but widely used eq. \ref{eq:tb_lin}. We see that the estimated spin temperature using the linear equation is overestimated throughout the $T_b$ range considered. The difference is more at lower values of $T_b$ which is more clear in the bottom panel of Fig. \ref{fig:tb2ts-mean17}.

The measured $T_b$ by the EDGES experiment at redshift $z=17.2$ is in the range between $-1 \, {\rm K}$ and $-0.3 \, {\rm K}$ with $1 \sigma$ confidence and the mean is $\sim -0.5 \, {\rm K}$. Using eq. \ref{eq:tb} and assuming the mean HI density, the estimated accurate spin temperatures for $T_b=-1$, $-0.5$ and $-0.3 \, {\rm K}$ at redshift $z=17.2$ are $1.38 \, {\rm K}$, $3 \, {\rm K}$ and $5 \, {\rm K}$ respectively. However, they are $1.72 \, {\rm K}$, $3.32 \, {\rm K}$ and $5.3 \, {\rm K}$ respectively which are $25\%$, $10.5\%$ and $6\%$ higher if the linear equation is used. This difference arises because the HI 21-cm optical depth $\tau_{b}$ becomes reasonably higher when $T_b$ is lower and, as a consequence, spin temperature $T_s$ becomes lower. The upper panel of Fig. \ref{fig:tb2ts-mean17}
shows the HI 21-cm optical depth $\tau_{b}$ for the same range of $T_b$ considered here. We see that $\tau_{b}$ rises to as high as $0.5$ for $T_b=-1 \, {\rm K}$ at redshift $z=17.2$. The linear approximation made in eq. \ref{eq:tb_lin} breaks down for higher $\tau_{b}$ values.  
In the {\it standard} scenario, where $T_b$ is expected to be around $-0.2$ K at redshift $z\sim 17$, the inferred spin temperature is $7.26$ K if we use the exact relation (eq. \ref{eq:tb}) which is $4\%$ lower compared to the spin temperature inferred using the linear equation (eq. \ref{eq:tb_lin}). 

The IGM kinetic temperature would be the same as the spin temperature estimated above under the assumption that they are tightly coupled with each other  through Ly-$\alpha$ coupling during cosmic dawn. Therefore, it is important that we use the exact equation in order to estimate the spin and IGM kinetic temperature accurately. 

Cold IGM during cosmic dawn not only alters the mean $T_{b}$, but also impacts fluctuations in the HI differential temperature $T_b$ which, in turn, impacts the statistical quantities such as the variance, skewness, kurtosis, power spectrum estimations. The following section investigates these using numerical simulations.


\section{Impact of high optical depth on HI fluctuations}
\label{sec:sim}
Apart from the mean HI differential brightness temperature high optical depth has impacts on the fluctuations in the differential brightness temperature and, therefore, on statistical quantities such as the variance, skewness, kurtosis, power spectrum, etc. We use numerical simulations to study the impact of high optical depth on the above statistical quantities.

\subsection{Scenarios \& phases of Cosmic Dawn}

We consider three different at redshifts $19.2$, $17.3$, and $15.4$ while simulating $T_b$ maps for the analysis carried out in this work. These three redshifts correspond to three distinct stages of the EDGES HI 21-cm absorption profile measured at the low band. At redshift $z=17.3$, the IGM kinetic temperature is fully coupled to the spin temperature but the heating of IGM is yet to start. This assumption makes $\TS$ equivalent to $\TK$ at that stage of Cosmic Dawn. The amplitude of the HI differential temperature is maximum here. At $z=19.2$, the coupling between the IGM kinetic and spin temperature is not fully complete, and this causes significant fluctuations in the spatial distribution of spin temperature. On the other hand, Ly-$\alpha$ coupling is fully complete at redshift $z=15.4$. As the heating starts at $z=17.3$, it affects the brightness temperature at redshift $z=15.4$.  The heating rate is more near to the sources and this causes fluctuations in the spin temperature distribution. 

We simulate $T_b$ maps for two different possible scenarios of cosmic dawn. In the first, the IGM temperature evolves as per the standard scenario i.e., there is no excess cooling or radio background. Accordingly, the IGM temperature is $\sim  7$ K at redshift $z=17.3$. This results in a signal with $\TB\approx 0.2$ K which is expected in the standard calculations \citep{pablo20, xu21}. At higher redshifts the IGM temperature scales as $\sim (1+z)^2$. The X-ray heating of the IGM starts at redshift $z = 17.3$. No IGM heating is considered at redshifts $z>17.3$. We denote this as `{\it standard}' scenario.

In the second scenario, we  assume that the mean HI differential brightness temperature $\TB$ at $z=19.2$, $17.3$ and $15.4$ are $\approx -0.35, -0.53, -0.3$ K respectively. These are in agreement with the EDGES low band measurements. We also assume that the IGM is significantly colder than the standard scenario predictions. We assume that $T_k=2.8$ K in all neutral voxels of simulation  at redshift $z=17.3$ in order to mimic EDGES measurements. Additionally, we assume that $T_k$ scales as $ (1+z)^2$ at higher redshifts. Like the first scenario, the X-ray heating in this scenario also starts at redshift $z \sim 17.3$ and we do not consider any IGM heating beyond $z>17.3$. The heating model parameters are tuned to produce the mean $T_b$ that is consistent with the EDGES measurements. We denote this as `{\it excess-cooling}’ scenario. We list various parameters related to these two scenarios in Table. \ref{table1}.


\begin{figure*}
	\includegraphics[width=\textwidth]{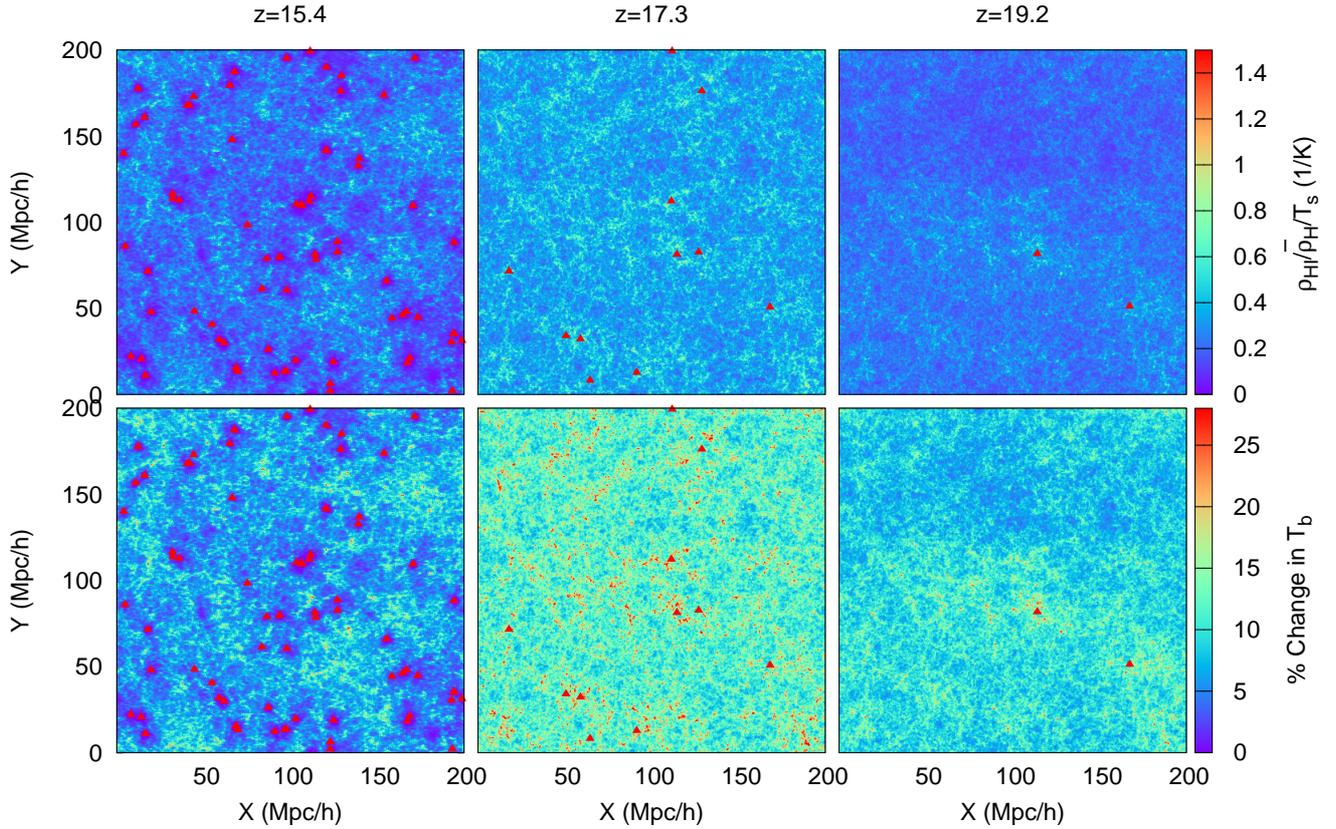}
    \caption{Top panels show $\frac{1}{T_{\rm S}}\frac{\rho_{\rm HI}}{\overline{\rho_{\rm H}}}$ of a randomly chosen slice from simulation at three different redshifts of Cosmic Dawn. Bottom panels show the `\% change' in $\TB$ (i.e., $\frac{\TBLIN-\TB}{\TB}\times 100$) due to the use of the linearized  form of the brightness temperature. These slices represent the {\it excess-cooling} scenario. Red solid triangles indicate locations of luminous sources.}
    \label{fig:dTB_slicezall}
\end{figure*}

\begin{table}
\begin{center}
\begin{tabular}{|l||*{3}{c|}}\hline
\backslashbox{IGM}{Redshift}
&\makebox[3.6em]{15.4}&\makebox[3.6em]{17.3}&\makebox[3.6em]{19.2}\\\hline\hline
Ly$\alpha$-coupling &Saturated&Saturated&Fluctuations\\\hline
X-ray heating &Yes&No&No\\\hline
Ionization &No&No&No\\\hline
$\TK(z)$-cold {\it Excess-cooling} & 2.22 K & 2.8 K & 3.45 K \\\hline
$\TK(z)$-cold {\it Standard} & 5.55 K & 7 K & 8.62 K \\\hline
$\TB(z)$-{\it Standard} & -0.19 K & -0.21 K & -0.13 K \\\hline
$\TB(z)$-{\it Excess-cooling} & -0.36 K & -0.53 K & -0.36 K \\\hline
\end{tabular}
\caption{Different simulated Cosmic Dawn scenarios at three different redshifts as considered in this study. }
\label{table1}
\end{center}
\end{table}

\subsection{Simulation of 21-cm signal maps}
We simulate the above-mentioned scenarios at redshifts $19.2$, $17.3$ and $15.4$. We use the {\sc grizzly} code \citep{ghara15a, ghara18} to simulate 21-cm signal maps for the redshifts of our interest. {\sc grizzly} requires cosmological density fields on grids, halo catalogs and source models as inputs. Dark matter density fields were generated using  the $N$-body  code {\sc cubep}$^3${\sc m}\footnote{\tt http://wiki.cita.utoronto.ca/mediawiki/index.php/CubePM} \citep{Harnois12}. The details of the simulation can be found in \citet{ghara15b}. The density fields were generated in a box of comoving size 200~$h^{-1}$ Mpc, which were further smoothed onto $432^3$ grids. We assume that the baryonic density fields follow the dark matter fields. The dark matter halos are identified using an on-the-fly halo finder which uses the spherical overdensity method. The mass resolution of the simulation (mass of dark matter particles is $1 \times 10^8\,  \MSUN$~$h^{-1}$) limits in resolving the halos with mass $\lesssim 5 \times 10^9\, \MSUN$~$h^{-1}$.

The {\sc grizzly} code is based on a one-dimensional radiative transfer scheme and is an independent implementation of the previously developed {\sc bears} algorithm  \citep{Thom09,2018NewA...64....9K}. Apart from the density fields and halo catalog at the redshifts of our interest, it also needs a source model to simulate HI differential temperature maps in the simulation box. We assume all the halos which are identified using the halo finder algorithm host sources of radiation. We also assume that the stellar mass ($M_\star$) associated with a halo is proportional to the mass of the halo ($M_{\rm halo}$), i.e., $M_\star = f_\star\times \frac{\OmegaB}{\Omegam} M_{\rm halo}$. The value of the star formation efficiency $f_\star$ is uncertain during the Cosmic Dawn, we assume a fixed  star formation efficiency $f_\star=0.1$ for our study.

We expect the IGM to be mostly neutral at the epochs of our interest. Therefore, we ignore the ionization fraction calculation and assume the IGM to be completely neutral at these three redshifts. For the {\it excess-cooling} scenario at $z\approx 19.2$, we assume the background IGM temperature to be $3.45$ K as discussed above. Further, we tune the rate of emission of $\lya$ photons per stellar mass so that the resulting average $\TB$ is  $-0.35$ K similar to the measured value by the EDGES at that redshift. 
We use the same source properties for the {\it standard} scenario at $z=19.2$. The only difference is that the IGM temperature is assumed to be $8.62$ K for the {\it standard} scenario.  

For the HI 21-cm signal at redshift $z=17.3$, the simulated 21-cm signal depends only on the density field at that redshift and the background IGM kinetic temperature as we assume $\lya$ saturated IGM. As discussed before, we set the IGM  spin temperature in each neutral voxel of simulation to a constant $7$ K and $2.8$ K at $z= 17.3$ in the {\it standard } and {\it excess-cooling}  scenarios respectively. Ideally spin temperature should depend on the local HI density due to adiabatic expansion of HI gas. This will cause the spin temperature to vary around the mean spin temperature. However, we ignore such effect here and defer the issue for future study. 

For the 21-cm signal at redshift $z = 15.4$, we assume that parts of the IGM around the sources are heated by X-rays radiation emitted from the sources. The IGM kinetic temperature is highly coupled to the spin temperature through Ly-$\alpha$ coupling i.e.,  $\TS=\TK$. For the {\it excess-cooling} scenario, we tune the emissivity of the X-ray photons per unit stellar mass such that the average differential brightness temperature  matches with the one measured by the EDGES-low.
The spectrum of an X-ray source is assumed to follow a power-law with a spectral index of $1.2$ which roughly represents a Quasar type source. We use the same X-ray source model for the {\it standard} scenario and find that the  mean $\TB$ is $-0.19$ K at that redshift.

Based on the prescription described above, we calculate Ly-$\alpha$ coupling coefficients and IGM kinetic temperature on each grid of simulated boxes at redshifts $z=19.2$ and $15.4$ respectively. We then generate the spin temperature $T_s$ for all three redshifts and HI 21-cm optical depth $\tau_b$ maps using formulae described in section \ref{sec:global}. Finally, we simulate two separate differential brightness temperature maps $\TB(\mathbf{x}, z)$ using the exact equation \ref{eq:tb}
and linearized equation \ref{eq:tb_lin}. We do it for the {\it standard} and {\it excess-cooling} scenarios separately. 

\section{Results}
\label{sec:res}
In this section, we describe our results on the impact of high optical depth on $T_b$ images and statistical quantities such as the mean, variance, skewness, kurtosis, and power spectrum of HI differential brightness temperature. We also present equations that help us understand the results on the power spectrum. 

\subsection{Impact on images}
Fig. \ref{fig:dTB_slicezall} demonstrates impact of high optical depth $\tau_b$ on the differential brightness temperature images.  The top panels, from left to right, show the quantity $\frac{1}{T_s}\frac{\rho_{\rm HI}}{\overline{\rho}_{\rm H}}$ which is a scaled version of the HI 21-cm optical depth $\tau_b$ at redshifts $z=15.4$, $17.3$ and $19.2$ respectively.  $\tau_b$ traces the underlying HI density field at redshift $z=17.3$ (Top middle panel) as the spin temperature is assumed to be uniform at $2.8$ K. We find that the mean $\tau_b$ is $0.2$, and there are a large number of grid points which have higher optical depth than the mean. We then estimate the differential brightness temperature maps using the exact (eq. \ref{eq:tb}) and linear relations (eq. \ref{eq:tb_lin}). The bottom panels show the percentage change in $\TB$, (i.e. ($T_{b-{\rm lin}}-\TB)/\TB \times 100$) between the two maps. We see that the amplitude of the differential brightness temperature is overestimated when the linear equation is used. The overestimation is more for regions with higher density and can be as high as $\sim 30\%$.  This is expected as the optical depth is higher in higher density regions and, thus, modification to the HI differential brightness temperature is more.

\begin{figure}
	\includegraphics[width=\columnwidth]{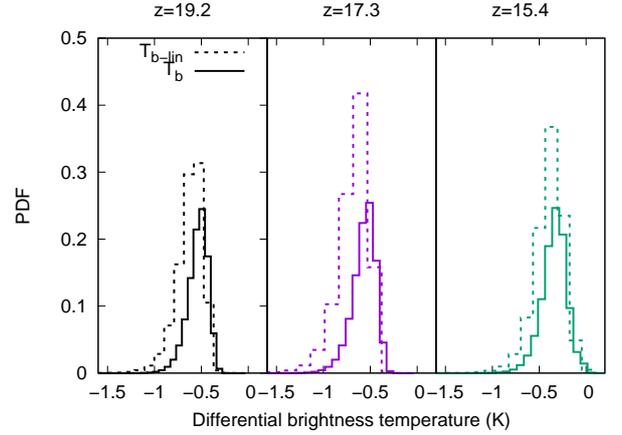}
    \caption{This figure compares  the probability distribution function (PDF) of the differential temperature estimated using the exact (Eq. \eqref{eq:tb}) and linearized equations Eq. \eqref{eq:tb_lin})  at three different redshifts for the {\it excess-cooling} scenario. }
    \label{fig:PDFtbalt}
\end{figure}

The impact at redshift $z=15.4$ is slightly different. Here, the kinetic and spin temperature around sources (marked with the filled triangles) in Fig. \ref{fig:dTB_slicezall} is very high due to X-ray heating. Therefore, the HI optical depth is very low. However, far from sources X-ray heating is not efficient and the HI density is also modest. Consequently, the HI optical depths in these regions are high. Regions that are very far from sources are mostly voids which again cause the optical depth to remain very low. Therefore, the impact due to high optical depth on the HI differential brightness temperature is the strongest in the intermediate regions where X-ray heating is not sufficient and HI density is moderate. This can be seen from the bottom left panel which shows that changes in $T_b$ near and far to sources are minimum. At redshift $z=19.2$, the spin temperature is strongly coupled to the background IGM kinetic temperature in the vicinity of Ly-$\alpha$ sources, but it is coupled to the CMBR temperature far from the sources. Consequently, the spin temperature is very low near the source. Therefore, the effect of high optical depth is strong near to sources, but not so much far from sources. Overall, we find that $\TB$  at redshifts $z=15.4$ and $19.2$ is over-predicted by $10-20\%$ which is less compared to what we find at redshift $17.3$. This is mainly because the effective spin temperature is higher at these two redshifts compared to that at $z=17.3$.  

Fig. \ref{fig:PDFtbalt} compares the probability distribution function (PDF) of the differential temperature estimated using the exact and linear methods. We note that $T_b$ with large negative values are suppressed significantly when the exact method is used. $T_b$ with large negative values are associated with high-density regions and low spin temperature and, therefore, with large HI 21-cm  optical depth. This causes a significant suppression in the $T_b$ values. Further, it appears that asymmetricity and width of the PDFs are reduced when $T_b$ is calculated using the exact equation.  

The results shown above are typical for the {\it excess-cooling} scenario. However, similar qualitative results are seen for the {\it standard} scenario which we do not show here explicitly.

\begin{figure}
	\includegraphics[width=\columnwidth]{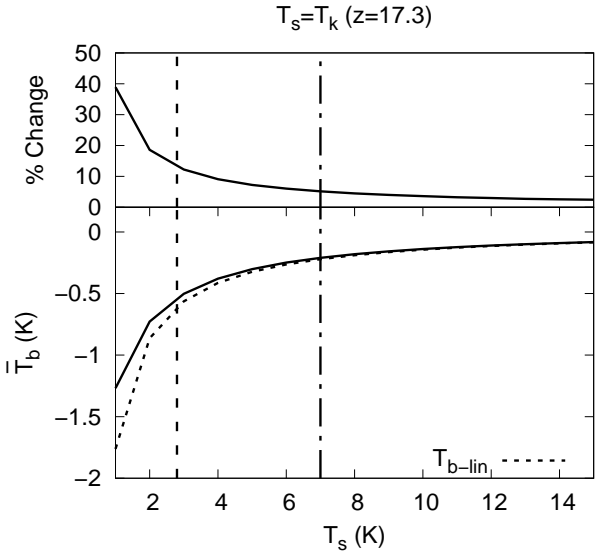}
    \caption{The bottom panel shows the average brightness temperature estimated from the simulation box as a function of IGM temperature at $z=17.3$. The solid curve represents the use of the exact form of the brightness temperature, while the dashed curve is for the approximate linear form of the same. We assume strong Ly-$\alpha$ coupling and the spin temperature to be constant over the IGM. The upper panel shows the \% change in the brightness temperature. The vertical dashed and dashed-dotted lines indicate gas temperature considered for the {\it excess-cooling} and {\it standard} scenarios considered in this study.}
    \label{fig:dTB_TK_TSz17}
\end{figure}

\subsection{Impact on the global 21-cm signal}
Using eq. \ref{eq:tb-nonlin} the fractional change in the estimation of the mean $T_b$ for uniform $T_s$ background can be better approximated as, \begin{equation}
\frac{\overline{T}_{b-\rm lin}  -\overline{T}_{b}}{ \overline{T}_{b}} \approx \frac{\bar{\tau}_b C_2}{2 -\bar{\tau}_b C_2}, 
\label{eq:tb-change}
\end{equation}
where $\bar{\tau}_b (z)=\frac{4 \, {\rm mK}}{T_s(z)} \left (\frac{\Omega_{b0}h^2}{0.02} \right) \left (\frac{0.7}{h} \right ) \left(\frac{H_0}{H(z)}\right) (1+z)^3 $ and 
$C_2=\overline{ \rho^2_{\rm HI}}/\overline{ \rho_{\rm HI} }^2$. We find $C_2 \approx 1.075624$ at $z=17.3$ calculated at scale $0.46 h^{-1}$ Mpc which is the resolution of our simulation. For perfectly uniform density distribution $C_2=1$. However, due to fluctuations in density distribution, $C_2$ becomes higher than one. It should, in principle, depend on the resolution of the simulation. We see from eq. \ref{eq:tb-change} that the percentage change in the mean differential brightness temperature scales  $\sim 1/T_s$ i.e., for higher background spin temperature the effect is less. This is also seen in the top panel of Fig. \ref{fig:dTB_TK_TSz17}.  Nonetheless, from simulation, we find that the linear equation over predicts the mean differential brightness temperature at $z=17.3$ by $\sim 5\%$  for the {\it standard} scenario, whereas is it $16 \%$  in the {\it excess-cooling} scenario. Eq. \ref{eq:tb-change} gives similar numbers for both scenarios.

The amount of overestimation of the mean differential brightness temperature at $z=17.3$ by the linear equation increases for a colder IGM.  The bottom panel of Fig. \ref{fig:dTB_TK_TSz17} shows the mean $T_b$ calculated using the exact and linear forms as a function of spin temperature. The spin temperature is the same as the background IGM temperature $\TK$ at redshift $17.3$. The top panel plots the percentage change in $\TB$ due to the usage of the linear form. The change in the mean signal increases rapidly with the decrease of $\TS$, while it is negligible for $\TS \gtrsim 10$ K. 

It is not straightforward to obtain an analytic form for the percentage change in the mean signal for redshifts $15.4$ and $19.2$ since spin temperature fluctuates at both these redshifts. Nonetheless, from simulations we find that the mean brightness temperature is overpredicted by $\sim 4\%$ both at redshift $15.4$ and $19.2$ in the {\it standard} scenario if it is calculated using the linear equation. For the {\it excess-cooling} scenario the mean brightness temperature is overpredicted by $\sim 8\%$ and $10\%$ at redshift $15.4$ and $19.2$ respectively. The above results are summarized in Table \ref{table1} and in Fig. \ref{fig:stats}. 

We note that the results discussed above are expected to change if simulations with different resolutions are used. The mean differential brightness temperature will be suppressed more in comparison to the linear predictions for high-resolution simulations. This is because the density clumpiness which is quantified through the quantity $C_2$ will increase. Here, we focus on the impact of high optical depth and defer the discussion on the impact of density clumpiness for future works.

\subsection{Impact on the one point statistics}
\begin{figure*}
	\includegraphics[width=\textwidth]{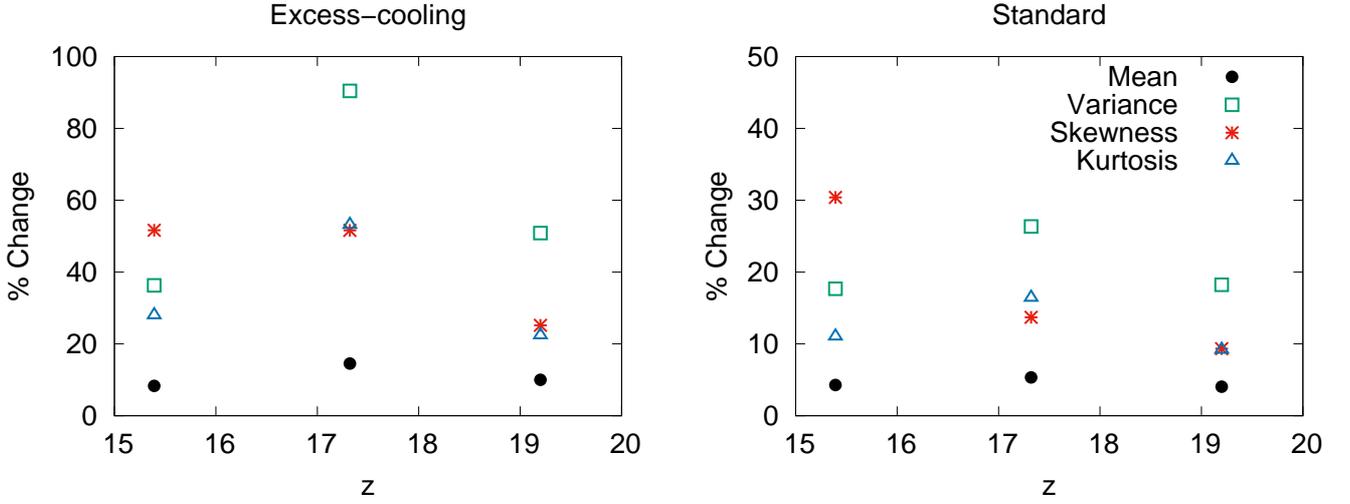}
    \caption{The figure shows  \% change in the estimation of the mean and various one point statistics of the HI differential brightness temperature when the linearized equation (eq. \ref{eq:tb_lin}) is used.  The Left and right panels represent the {\it excess-cooling} and {\it standard} scenarios considered in this study. Note that all these quantities are estimated at the simulation resolution. }
    \label{fig:stats}
\end{figure*}

We find that (in Fig. \ref{fig:stats} ) the variance of HI differential brightness fluctuations calculated at the simulation resolution is over-predicted by $\sim 18\%$, $\sim 26 \%$ and $\sim 18 \%$ at redshift $z=15.4$, $17.3$ and $19.2$ respectively for the {\it standard} scenario when the linearized equation is used. In the {\it excess-cooling} scenario the corresponding changes are as large as $\sim 36 \%$, $90\%$ and $50\%$ respectively. This can be explained from Fig. \ref{fig:PDFtbalt}. We see that the probability distribution function (PDF) is squeezed when the exact equation is used to estimate HI differential brightness. In particular, differential brightness temperatures with large negative values are shifted towards less negative values. Because, those points belong to high HI density and low spin temperature i.e., are associated with high HI optical depth. Therefore,  the amplitude of $T_b$ is reduced significantly when the exact equation is used in comparison to the predictions from the linear equation.

\begin{figure*}
	\includegraphics[width=\textwidth]{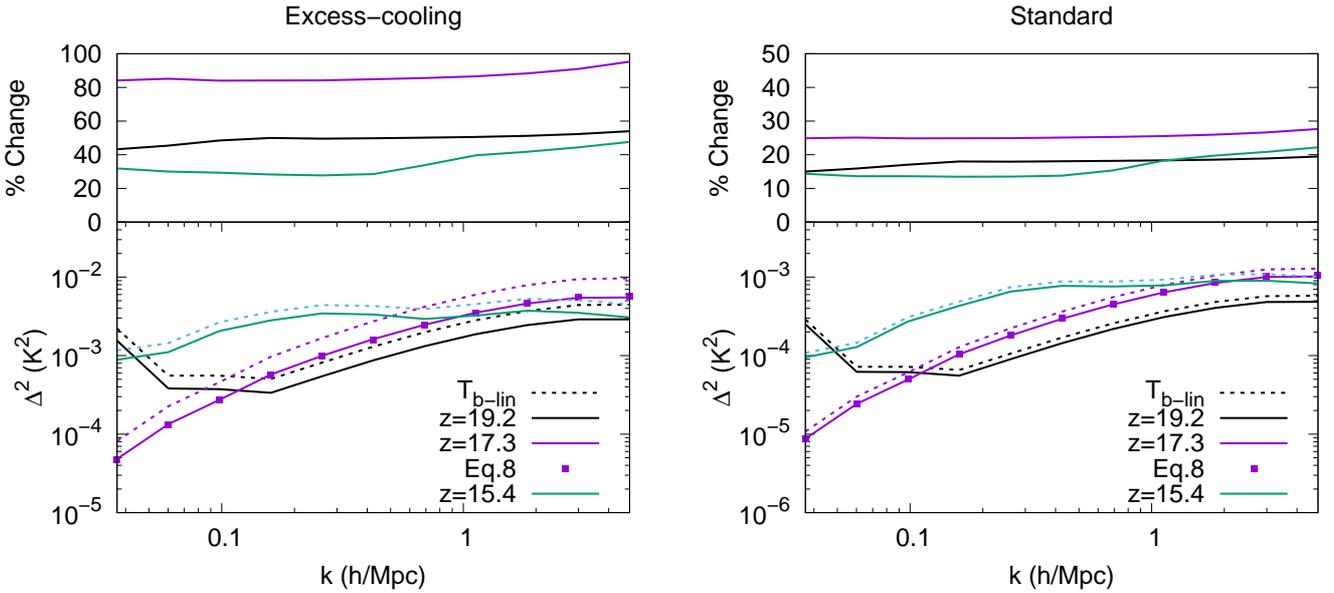}
    \caption{Bottom panel: The dimensionless spherically averaged power spectrum of the HI differential brightness temperature $T_b$ at three different stages of Cosmic Dawn. The solid (dashed) curves represent the simulated power spectra estimated using the exact (linear) form of the brightness temperature equation.  The top panel shows the \% difference in the power spectra between the two. The left and right panels represent the {\it excess-cooling} and {\it standard} scenarios considered in this study. The filled squares in bottom panels are improved estimates of  power spectra using analytical equation \ref{eq:pk_exact_ana}.}
    \label{fig:PS}
\end{figure*}

Next we focus on the higher order moments of $T_b$ i.e., skewness and kurtosis  which are defined as,
\begin{equation}
    \mu_i = \left\langle \left\{ \frac{T_b({\bf x}, z) -\bar{T}_b(z)}{\sigma_{T_b}} \right\}^i \right\rangle,
\end{equation}
where $i=3$ and $4$ correspond to skewness and kurtosis respectively. $\sigma_{T_b}$ is the standard deviation. We use simulations and find (shown in Fig. \ref{fig:stats} ) that the linear equations over-predicts the skewness by $\sim 30\%$, $14 \%$ and $9 \%$ at redshift $z=15.4$, $17.3$ and $19.2$ respectively for the {\it standard} scenario.  In the {\it Excess-cooling} scenario the corresponding changes are $\sim 50 \%$, $50\%$ and $25\%$ respectively.

  The skewness is a measure of lack of symmetry around the mean of the distribution of differential brightness temperature. The skewness for a normal distribution is zero. Distributions of the HI overdensity and spin temperature deviate from the normal distribution, and thus, introduce certain levels of skewness.  In the linear approximation,  the PDF of $\TB$  follows the PDF of HI overdensity. The use of the exact form (eq. \ref{eq:tb-nonlin}) reduces the amplitude of the absorption signal. This effect is more significant at the denser regions with lower spin temperature (see Fig. \ref{fig:PDFtbalt}). This makes the PDF of $\TB$ more symmetric for the exact calculation of $\TB$, and reduces the skewness by $\sim 15 \%$ and $\sim 50 \%$ at redshift $z=17.3$ for the {\it standard} and {\it excess-cooling} scenarios respectively. Similar results can also be found  for  $z=19.2$ (see the left panel of Fig. \ref{fig:PDFtbalt}). We see that the value of the percentage change is quite large for redshift $15.4$. For this redshift, the IGM  consists of regions both with emission and absorption signals. In presence of the X-ray heating, the denser regions around the sources are heated which turns the signal into emission. The effect of using the exact form is negligible for these emission regions, while more significant for the tail of the PDF with large amplitude. This makes the PDF more symmetric and thus produces a larger value for the change.

  The kurtosis measures outliers of a distribution. It determines whether the distribution is light-tailed or heavy-tailed relative to a normal distribution. The use of the exact form of brightness temperature reduces the kurtosis of the $\TB$ fields. We find that the linear equation of brightness temperature over-predicts the kurtosis by $\sim 30\% (11\%)$, $50\% (16\%)$ and $20\% (9\%)$ at redshift $z=15.4$, $17.3$ and $19.2$ respectively for the {\it excess-cooling} ({\it standard}) scenario. 
  
  Like the mean differential temperature, the impact of high optical depth on various one point statistics will change if calculated at different scales. The effects are expected to get reduced/enhanced at larger/smaller scales in compare to the scale ($\sim 0.5h^{-1}\, {\rm Mpc^{-1}} $) discussed above.

\subsection{Impact on the HI 21-cm power spectrum}
Here we focus on the spherically averaged redshifted HI 21-cm power spectrum. From simulations, we find that the HI power spectrum of differential brightness temperature calculated using the linear equation is overestimated by $\sim 25\%$ and $\sim 80\%$ for the {\it standard} and {\it excess-cooling} scenarios respectively at redshift $\approx 17.3$ (see the upper panels of Fig.\ref{fig:PS}).  At redshift $z=15.4$ and $19.2$ the corresponding changes  are found to be $\sim 15\%$ and $\sim 40 \%$ for the {\it standard} and {\it excess-cooling} scenarios respectively. We find that the change is nearly constant at  all scales we explore. The power spectrum is reduced when the exact equation are used, mainly because the HI fluctuations are reduced, particularly at high HI density and low spin temperature regions which have large HI optical depth.

Under the assumption of uniform spin temperature background, the redshifted HI 21-cm power spectrum in the linear case can be calculated as,
\begin{equation}
    P_{\rm HI-lin}(k, z) \simeq \left(\frac{T_s-T_{\gamma}}{1+z} \bar{\tau}_b \right)^2 P(k, z),
\end{equation}
where $P(k, z)$ is the dark matter power spectrum at redshift $z$, and hydrogen is assumed to be fully neutral and trace the underlying dark matter. However, using eq. \ref{eq:tb-nonlin} the redshifted HI 21-cm power spectrum can be better approximated as,
\begin{equation}
    P_{\rm HI-b}(k, z) \approx\left[\frac{T_s-T_{\gamma}}{1+z} (\bar{\tau}_b- \bar{\tau}^2_b) \right]^2 P(k, z).
    \label{eq:pk_exact_ana}
\end{equation}
 According to the above equations the fractional change in the HI power spectrum is $\approx \frac{2\bar{\tau}_b-\bar{\tau}^2_b}{(1-\bar{\tau}_b)^2}$ which predicts the changes  of $\sim 22\%$ and $\sim 71\%$ for the {\it standard} and {\it excess-cooling} scenarios respectively at $z=17.3$. This shows that the analytical expressions above can be used to correctly estimate the HI Power spectrum at any redshift for uniform spin temperature background. These equations are also useful in estimating errors that the linear equation makes in calculating the power spectrum for any uniform spin temperature background.

\section{Summary \& Discussion}
\label{sec:con}
The kinetic and spin temperature of HI in the IGM are likely to be the lowest just before the heating starts during cosmic dawn. Consequently, the HI 21-cm optical depth, which is inversely proportional to the spin temperature, becomes significantly large. In the {\it standard} scenario the mean HI 21-cm optical depth can be as large as $\sim 0.1$ during cosmic dawn.  The mean optical depth rises to $\sim 0.2$ if the IGM is significantly colder around $\sim 3$ K at redshift $z \sim 17$ which is a  possibility suggested by the EDGES low band observations. The HI 21-cm optical depth can be even higher at regions with higher HI density. 

We investigate the impact of the large HI optical depth on the redshifted HI 21-cm signal from cosmic dawn. We revisit the validity of the widely used equation (eq. \ref{eq:tb_lin}) for estimating the HI 21-cm differential brightness temperature.  The equation inherently assumes that the HI optical depth $\tau_b$ is much smaller than the unity and approximates the quantity $[1-\exp(-\tau_b)]$ as $\tau_b$. This equation is referred as linearized equation. Here, we consider two different scenarios, one without any additional cooling mechanism for the gas in the IGM ({\it standard} scenario) and the other with gas in the IGM much colder than expected (referred as {\it excess-cooling} scenario) for our study.  The second scenario is motivated by the EDGES observations which measured unusually strong HI 21-cm absorption signal.

In the first part, we focus on the  spin and IGM temperature inferred from an observed global HI 21-cm signal from cosmic dawn. We find that the linearized equation can overestimate the spin temperature by $\sim 5\%$ during the cosmic dawn in the {\it standard} scenario. For the EDGES like absorption profile and {\it excess-cooling} scenario the inferred spin and IGM kinetic temperature can be up by $\sim 10 \%$. The difference is more for lower HI differential brightness temperature. 

In the second part, we focus on the impact of large HI 21-cm optical depth on the statistics of the differential brightness temperature. We use numerical simulations and consider the above two scenarios i.e., the {\it standard} and {\it excess-cooling} for our study.  We find that the mean estimated differential brightness temperature is overestimated up to $\sim 5\%$ and $16 \%$ in the {\it standard} and {\it excess-cooling} respectively if the linearized equation is used.  However, we expect to find even more significant impact if simulations with finer resolution is used. The amount of overestimation  is even more at regions with higher HI density, which have relatively higher HI optical depth. As a results, the probability distribution function of differential brightness temperature is squeezed and looks more like a Gaussian if the exact equation (eq. \ref{eq:tau}) is used. This is also evident from the analysis of variance and skewness measurements. Both the quantities are reduced by a significant fraction if the exact equation used. For example, the variance calculated at simulation resolution is reduced up to $\sim 30\%$ and $\sim 90\%$ in the  {\it standard} and {\it excess-cooling} scenarios respectively if calculated properly. We find similar effects on the skewness and kurtosis. We also notice a remarkable change in the skewness at redshift $z=15.4$ although the effective spin temperature/optical depth is higher/lower in compared to those at redshift $z=17.3$. This is mainly due to the fact that IGM temperature and spin temperature near to sources are very high, but are lower far from sources. In presence of the X-ray heating, the denser regions around the sources are heated above the CMBR temperature which turns the HI 21-cm signal into emission. This makes the probability distribution  of the differential brtightness temperature more symmetric in compare to that at higher redshifts. We expect the above effect to dilute to some extent at larger scales as over density regions will be smoothed out. A detailed investigation is needed to assess the impact of the high optical depth on the one point statistics which is deferred for future work.  

We also look at the impact on the spherically averaged HI 21-cm power spectrum. We find that the power spectrum estimated from the linearized differential brightness temperature is over predicted by $\sim 25\%$ and $\sim 80 \%$ for the {\it standard} and {\it excess-cooling}  scenarios respectively just before the IGM heating starts and when the spin temperature is assumed to be uniformly distributed. The change is found to be the same in all scales, except at very small scales where the change is higher. The power spectrum is over predicted by $\sim 15\%$ and $\sim 40\%$ in the {\it standard} and {\it excess-cooling}  scenarios respectively both at redshifts $z=15.4$ and $19.2$ when the effective spin temperature is higher compared to that  at $z=17.3$. We present equations which can used to analytically estimate the HI 21-cm power spectrum for uniform spin temperature background at any redshift more accurately in moderately large HI optical depth. We show that predictions from these equations are in agreement with results from simulations.

At the end we note that the effect demonstrated here is quite generic as it arises due to low spin temperature and presence of HI over density regions during cosmic dawn. However, we expect our results to change to some extent for different source models. For example, IGM heating will be more uniform if sources with hard X-ray spectra (e.g. high mass X-ray binaries) are considered \citep{fialkov14a, 2019MNRAS.487.2785I}. Given the same differential brightness temperature $T_b$ at $z=15.4$, the change in the power spectra due to properly accounting the high HI optical depth will be around $\sim 27\%$ if the kinetic and spin temperature are uniformly distributed. This change is slightly lower compared to the case when $T_s$ is in-homogeneously distributed as discussed above. In another situation, if the formation of the first generation of luminous sources is delayed, universe will get more time to cool down adiabatically. In that case, the background IGM and spin temperature  will be lower which will result in higher optical depth and the effects on the statistical quantities of the HI differential temperature will be more prominent. Apart from that, our assumption of a uniform spin temperature background at redshift $z=17.3$ may not be realistic. The IGM kinetic temperature of HI, which is expanding adiabatically, scales as $\sim \rho^{2/3}_{\rm HI}$ \citep{xu21}. Incorporating such density dependence temperature distribution may affect our results at $z=17.3$. Another assumption that the kinetic temperature is strongly coupled to the spin temperature at redshifts $z \lesssim 17.3$ is not guaranteed by the 
EDGES absorption profile. For weak or moderate coupling, the inferred kinetic temperature will be lower than the one estimated here. Accurate estimation of the kinetic temperature will depend on the specific source model. Despite all such limitations, we believe that results presented here demonstrate, for the first time, the importance of properly accounting for the effect of high optical depth on the statistical quantities that the ongoing and upcoming experiments aim to measure. A detailed and through investigation is required to assess the impact of large HI optical depth on the 21-cm signal from cosmic dawn.

\section*{Acknowledgements}
KKD  acknowledges financial support from BRNS through a project grant (sanction no: 57/14/10/2019-BRNS). RG acknowledge support by the Israel Science Foundation (grant no. 255/18). We thank Somnath Bharadwaj and Tirthankar Roy Choudhury for useful discussions and Ankita Bera for her help in improving the quality of the manuscript.

\section*{Data Availability}

The data underlying this work will be shared upon reasonable request to the corresponding author.


\bibliographystyle{mnras}
\bibliography{mybib, RGbib} 

\begin{thebibliography}{}
\makeatletter
\relax
\def\mn@urlcharsother{\let\do\@makeother \do\$\do\&\do\#\do\^\do\_\do\%\do\~}
\def\mn@doi{\begingroup\mn@urlcharsother \@ifnextchar [ {\mn@doi@}
  {\mn@doi@[]}}
\def\mn@doi@[#1]#2{\def\@tempa{#1}\ifx\@tempa\@empty \href
  {http://dx.doi.org/#2} {doi:#2}\else \href {http://dx.doi.org/#2} {#1}\fi
  \endgroup}
\def\mn@eprint#1#2{\mn@eprint@#1:#2::\@nil}
\def\mn@eprint@arXiv#1{\href {http://arxiv.org/abs/#1} {{\tt arXiv:#1}}}
\def\mn@eprint@dblp#1{\href {http://dblp.uni-trier.de/rec/bibtex/#1.xml}
  {dblp:#1}}
\def\mn@eprint@#1:#2:#3:#4\@nil{\def\@tempa {#1}\def\@tempb {#2}\def\@tempc
  {#3}\ifx \@tempc \@empty \let \@tempc \@tempb \let \@tempb \@tempa \fi \ifx
  \@tempb \@empty \def\@tempb {arXiv}\fi \@ifundefined
  {mn@eprint@\@tempb}{\@tempb:\@tempc}{\expandafter \expandafter \csname
  mn@eprint@\@tempb\endcsname \expandafter{\@tempc}}}

\bibitem[\protect\citeauthoryear{{Barkana}}{{Barkana}}{2018}]{Barkana18Nature}
{Barkana} R.,  2018, \mn@doi [\nat] {10.1038/nature25791}, \href
  {https://ui.adsabs.harvard.edu/abs/2018Natur.555...71B} {555, 71}

\bibitem[\protect\citeauthoryear{{Barry} et~al.,}{{Barry}
  et~al.}{2019}]{barry19}
{Barry} N.,  et~al., 2019, \mn@doi [\apj] {10.3847/1538-4357/ab40a8}, \href
  {https://ui.adsabs.harvard.edu/abs/2019ApJ...884....1B} {884, 1}

\bibitem[\protect\citeauthoryear{{Bharadwaj} \& {Ali}}{{Bharadwaj} \&
  {Ali}}{2005}]{bharadwaj05}
{Bharadwaj} S.,  {Ali} S.~S.,  2005, \mnras, \href
  {http://adsabs.harvard.edu/abs/2005MNRAS.356.1519B} {356, 1519}

\bibitem[\protect\citeauthoryear{{Bowman} \& {Rogers}}{{Bowman} \&
  {Rogers}}{2010}]{bowman10}
{Bowman} J.~D.,  {Rogers} A.~E.~E.,  2010, \nat, \href
  {http://adsabs.harvard.edu/abs/2010Natur.468..796B} {468, 796}

\bibitem[\protect\citeauthoryear{{Bowman} et~al.,}{{Bowman}
  et~al.}{2013}]{bowman13}
{Bowman} J.~D.,  et~al., 2013, \mn@doi [\pasa] {10.1017/pas.2013.009}, \href
  {http://adsabs.harvard.edu/abs/2013PASA...30...31B} {30, 31}

\bibitem[\protect\citeauthoryear{{Bowman}, {Rogers}, {Monsalve}, {Mozdzen}  \&
  {Mahesh}}{{Bowman} et~al.}{2018}]{EDGES18}
{Bowman} J.~D.,  {Rogers} A. E.~E.,  {Monsalve} R.~A.,  {Mozdzen} T.~J.,
  {Mahesh} N.,  2018, \mn@doi [\nat] {10.1038/nature25792}, \href
  {https://ui.adsabs.harvard.edu/\#abs/2018Natur.555...67B} {555, 67}

\bibitem[\protect\citeauthoryear{{Chakraborty} et~al.,}{{Chakraborty}
  et~al.}{2021}]{chakraborty21}
{Chakraborty} A.,  et~al., 2021, \mn@doi [\apjl] {10.3847/2041-8213/abd17a},
  \href {https://ui.adsabs.harvard.edu/abs/2021ApJ...907L...7C} {907, L7}

\bibitem[\protect\citeauthoryear{{Choudhuri}, {Ghosh}, {Roy}, {Bharadwaj},
  {Intema}  \& {Ali}}{{Choudhuri} et~al.}{2020}]{choudhuri20}
{Choudhuri} S.,  {Ghosh} A.,  {Roy} N.,  {Bharadwaj} S.,  {Intema} H.~T.,
  {Ali} S.~S.,  2020, \mn@doi [\mnras] {10.1093/mnras/staa762}, \href
  {https://ui.adsabs.harvard.edu/abs/2020MNRAS.494.1936C} {494, 1936}

\bibitem[\protect\citeauthoryear{{Datta}, {Mellema}, {Mao}, {Iliev}, {Shapiro}
  \& {Ahn}}{{Datta} et~al.}{2012}]{datta12}
{Datta} K.~K.,  {Mellema} G.,  {Mao} Y.,  {Iliev} I.~T.,  {Shapiro} P.~R.,
  {Ahn} K.,  2012, \mn@doi [\mnras] {10.1111/j.1365-2966.2012.21293.x}, \href
  {http://adsabs.harvard.edu/abs/2012MNRAS.424.1877D} {424, 1877}

\bibitem[\protect\citeauthoryear{{Datta}, {Kundu}, {Paul}  \& {Bera}}{{Datta}
  et~al.}{2020}]{datta20}
{Datta} K.~K.,  {Kundu} A.,  {Paul} A.,   {Bera} A.,  2020, \mn@doi [\prd]
  {10.1103/PhysRevD.102.083502}, \href
  {https://ui.adsabs.harvard.edu/abs/2020PhRvD.102h3502D} {102, 083502}

\bibitem[\protect\citeauthoryear{{DeBoer} et~al.,}{{DeBoer}
  et~al.}{2017}]{2017PASP..129d5001D}
{DeBoer} D.~R.,  et~al., 2017, \mn@doi [Publications of the Astronomical
  Society of the Pacific] {10.1088/1538-3873/129/974/045001}, \href
  {https://ui.adsabs.harvard.edu/\#abs/2017PASP..129d5001D} {129, 045001}

\bibitem[\protect\citeauthoryear{{Fialkov}, {Barkana}  \& {Visbal}}{{Fialkov}
  et~al.}{2014}]{fialkov14a}
{Fialkov} A.,  {Barkana} R.,   {Visbal} E.,  2014, \mn@doi [\nat]
  {10.1038/nature12999}, \href
  {https://ui.adsabs.harvard.edu/abs/2014Natur.506..197F} {506, 197}

\bibitem[\protect\citeauthoryear{{Ghara} \& {Mellema}}{{Ghara} \&
  {Mellema}}{2020}]{2020MNRAS.492..634G}
{Ghara} R.,  {Mellema} G.,  2020, \mn@doi [\mnras] {10.1093/mnras/stz3513},
  \href {https://ui.adsabs.harvard.edu/abs/2020MNRAS.492..634G} {492, 634}

\bibitem[\protect\citeauthoryear{{Ghara}, {Choudhury}  \& {Datta}}{{Ghara}
  et~al.}{2015a}]{ghara15a}
{Ghara} R.,  {Choudhury} T.~R.,   {Datta} K.~K.,  2015a, \mn@doi [\mnras]
  {10.1093/mnras/stu2512}, \href
  {http://adsabs.harvard.edu/abs/2015MNRAS.447.1806G} {447, 1806}

\bibitem[\protect\citeauthoryear{{Ghara}, {Datta}  \& {Choudhury}}{{Ghara}
  et~al.}{2015b}]{ghara15b}
{Ghara} R.,  {Datta} K.~K.,   {Choudhury} T.~R.,  2015b, \mn@doi [\mnras]
  {10.1093/mnras/stv1855}, \href
  {http://adsabs.harvard.edu/abs/2015MNRAS.453.3143G} {453, 3143}

\bibitem[\protect\citeauthoryear{{Ghara}, {Choudhury}, {Datta}  \&
  {Choudhuri}}{{Ghara} et~al.}{2017}]{ghara16}
{Ghara} R.,  {Choudhury} T.~R.,  {Datta} K.~K.,   {Choudhuri} S.,  2017,
  \mn@doi [\mnras] {10.1093/mnras/stw2494}, \href
  {http://adsabs.harvard.edu/abs/2017MNRAS.464.2234G} {464, 2234}

\bibitem[\protect\citeauthoryear{{Ghara}, {Mellema}, {Giri}, {Choudhury},
  {Datta}  \& {Majumdar}}{{Ghara} et~al.}{2018}]{ghara18}
{Ghara} R.,  {Mellema} G.,  {Giri} S.~K.,  {Choudhury} T.~R.,  {Datta} K.~K.,
  {Majumdar} S.,  2018, \mn@doi [\mnras] {10.1093/mnras/sty314}, \href
  {http://adsabs.harvard.edu/abs/2018MNRAS.476.1741G} {476, 1741}

\bibitem[\protect\citeauthoryear{{Ghara} et~al.,}{{Ghara}
  et~al.}{2020}]{2020MNRAS.493.4728G}
{Ghara} R.,  et~al., 2020, \mn@doi [\mnras] {10.1093/mnras/staa487}, \href
  {https://ui.adsabs.harvard.edu/abs/2020MNRAS.493.4728G} {493, 4728}

\bibitem[\protect\citeauthoryear{{Ghara}, {Giri}, {Ciardi}, {Mellema}  \&
  {Zaroubi}}{{Ghara} et~al.}{2021}]{ghara2021}
{Ghara} R.,  {Giri} S.~K.,  {Ciardi} B.,  {Mellema} G.,   {Zaroubi} S.,  2021,
  \mn@doi [\mnras] {10.1093/mnras/stab776}, \href
  {https://ui.adsabs.harvard.edu/abs/2021MNRAS.503.4551G} {503, 4551}

\bibitem[\protect\citeauthoryear{{Greenhill} \& {Bernardi}}{{Greenhill} \&
  {Bernardi}}{2012}]{2012arXiv1201.1700G}
{Greenhill} L.~J.,  {Bernardi} G.,  2012, preprint, \href
  {http://adsabs.harvard.edu/abs/2012arXiv1201.1700G} {} (\mn@eprint {arXiv}
  {1201.1700})

\bibitem[\protect\citeauthoryear{{Greig} et~al.,}{{Greig}
  et~al.}{2021}]{2021MNRAS.501....1G}
{Greig} B.,  et~al., 2021, \mn@doi [\mnras] {10.1093/mnras/staa3593}, \href
  {https://ui.adsabs.harvard.edu/abs/2021MNRAS.501....1G} {501, 1}

\bibitem[\protect\citeauthoryear{{Harker} et~al.,}{{Harker}
  et~al.}{2009}]{harker2009}
{Harker} G. J.~A.,  et~al., 2009, \mn@doi [\mnras]
  {10.1111/j.1365-2966.2008.14209.x}, \href
  {https://ui.adsabs.harvard.edu/abs/2009MNRAS.393.1449H} {393, 1449}

\bibitem[\protect\citeauthoryear{{Harnois-D{\'e}raps}, {Pen}, {Iliev}, {Merz},
  {Emberson}  \& {Desjacques}}{{Harnois-D{\'e}raps} et~al.}{2013}]{Harnois12}
{Harnois-D{\'e}raps} J.,  {Pen} U.-L.,  {Iliev} I.~T.,  {Merz} H.,  {Emberson}
  J.~D.,   {Desjacques} V.,  2013, \mn@doi [\mnras] {10.1093/mnras/stt1591},
  \href {http://adsabs.harvard.edu/abs/2013MNRAS.436..540H} {436, 540}

\bibitem[\protect\citeauthoryear{{Islam}, {Ghara}, {Paul}, {Choudhury}  \&
  {Nath}}{{Islam} et~al.}{2019}]{2019MNRAS.487.2785I}
{Islam} N.,  {Ghara} R.,  {Paul} B.,  {Choudhury} T.~R.,   {Nath} B.~B.,  2019,
  \mn@doi [\mnras] {10.1093/mnras/stz1446}, \href
  {https://ui.adsabs.harvard.edu/abs/2019MNRAS.487.2785I} {487, 2785}

\bibitem[\protect\citeauthoryear{{Kamran}, {Ghara}, {Majumdar}, {Mondal},
  {Mellema}, {Bharadwaj}, {Pritchard}  \& {Iliev}}{{Kamran}
  et~al.}{2021}]{2021MNRAS.502.3800K}
{Kamran} M.,  {Ghara} R.,  {Majumdar} S.,  {Mondal} R.,  {Mellema} G.,
  {Bharadwaj} S.,  {Pritchard} J.~R.,   {Iliev} I.~T.,  2021, \mn@doi [\mnras]
  {10.1093/mnras/stab216}, \href
  {https://ui.adsabs.harvard.edu/abs/2021MNRAS.502.3800K} {502, 3800}

\bibitem[\protect\citeauthoryear{{Kapahtia}, {Chingangbam}, {Ghara}, {Appleby}
  \& {Choudhury}}{{Kapahtia} et~al.}{2021}]{2021JCAP...05..026K}
{Kapahtia} A.,  {Chingangbam} P.,  {Ghara} R.,  {Appleby} S.,   {Choudhury}
  T.~R.,  2021, \mn@doi [\jcap] {10.1088/1475-7516/2021/05/026}, \href
  {https://ui.adsabs.harvard.edu/abs/2021JCAP...05..026K} {2021, 026}

\bibitem[\protect\citeauthoryear{{Krause}, {Thomas}, {Zaroubi}  \&
  {Abdalla}}{{Krause} et~al.}{2018}]{2018NewA...64....9K}
{Krause} F.,  {Thomas} R.~M.,  {Zaroubi} S.,   {Abdalla} F.~B.,  2018, \mn@doi
  [\na] {10.1016/j.newast.2018.03.004}, \href
  {https://ui.adsabs.harvard.edu/abs/2018NewA...64....9K} {64, 9}

\bibitem[\protect\citeauthoryear{{Lewis} \& {Challinor}}{{Lewis} \&
  {Challinor}}{2007}]{lewis07}
{Lewis} A.,  {Challinor} A.,  2007, \mn@doi [\prd]
  {10.1103/PhysRevD.76.083005}, \href
  {https://ui.adsabs.harvard.edu/abs/2007PhRvD..76h3005L} {76, 083005}

\bibitem[\protect\citeauthoryear{{Majumdar}, {Pritchard}, {Mondal},
  {Watkinson}, {Bharadwaj}  \& {Mellema}}{{Majumdar} et~al.}{2018}]{majumdar18}
{Majumdar} S.,  {Pritchard} J.~R.,  {Mondal} R.,  {Watkinson} C.~A.,
  {Bharadwaj} S.,   {Mellema} G.,  2018, \mn@doi [\mnras]
  {10.1093/mnras/sty535}, \href
  {https://ui.adsabs.harvard.edu/abs/2018MNRAS.476.4007M} {476, 4007}

\bibitem[\protect\citeauthoryear{{Mellema}, {Koopmans}, {Shukla}, {Datta},
  {Mesinger}  \& {Majumdar}}{{Mellema} et~al.}{2015}]{2015aska.confE..10M}
{Mellema} G.,  {Koopmans} L.,  {Shukla} H.,  {Datta} K.~K.,  {Mesinger} A.,
  {Majumdar} S.,  2015, Advancing Astrophysics with the Square Kilometre Array
  (AASKA14), \href {http://adsabs.harvard.edu/abs/2015aska.confE..10M} {p.~10}

\bibitem[\protect\citeauthoryear{{Mertens} et~al.,}{{Mertens}
  et~al.}{2020}]{2020MNRAS.493.1662M}
{Mertens} F.~G.,  et~al., 2020, \mn@doi [\mnras] {10.1093/mnras/staa327}, \href
  {https://ui.adsabs.harvard.edu/abs/2020MNRAS.493.1662M} {493, 1662}

\bibitem[\protect\citeauthoryear{{Mondal} et~al.,}{{Mondal}
  et~al.}{2020}]{2020MNRAS.498.4178M}
{Mondal} R.,  et~al., 2020, \mn@doi [\mnras] {10.1093/mnras/staa2422}, \href
  {https://ui.adsabs.harvard.edu/abs/2020MNRAS.498.4178M} {498, 4178}

\bibitem[\protect\citeauthoryear{{Mu{\~n}oz}, {Dvorkin}  \& {Loeb}}{{Mu{\~n}oz}
  et~al.}{2018}]{Munoz18}
{Mu{\~n}oz} J.~B.,  {Dvorkin} C.,   {Loeb} A.,  2018, \mn@doi [\prl]
  {10.1103/PhysRevLett.121.121301}, \href
  {https://ui.adsabs.harvard.edu/abs/2018PhRvL.121l1301M} {121, 121301}

\bibitem[\protect\citeauthoryear{{Pal}, {Bharadwaj}, {Ghosh}  \&
  {Choudhuri}}{{Pal} et~al.}{2021}]{pal21}
{Pal} S.,  {Bharadwaj} S.,  {Ghosh} A.,   {Choudhuri} S.,  2021, \mn@doi
  [\mnras] {10.1093/mnras/staa3831}, \href
  {https://ui.adsabs.harvard.edu/abs/2021MNRAS.501.3378P} {501, 3378}

\bibitem[\protect\citeauthoryear{{Park}, {Mesinger}, {Greig}  \&
  {Gillet}}{{Park} et~al.}{2019}]{park19}
{Park} J.,  {Mesinger} A.,  {Greig} B.,   {Gillet} N.,  2019, \mn@doi [\mnras]
  {10.1093/mnras/stz032}, \href
  {https://ui.adsabs.harvard.edu/abs/2019MNRAS.484..933P} {484, 933}

\bibitem[\protect\citeauthoryear{{Patil} et~al.,}{{Patil}
  et~al.}{2014}]{patil2014}
{Patil} A.~H.,  et~al., 2014, \mn@doi [\mnras] {10.1093/mnras/stu1178}, \href
  {http://adsabs.harvard.edu/abs/2014MNRAS.443.1113P} {443, 1113}

\bibitem[\protect\citeauthoryear{{Patil} et~al.,}{{Patil}
  et~al.}{2017}]{patil17}
{Patil} A.~H.,  et~al., 2017, \mn@doi [\apj] {10.3847/1538-4357/aa63e7}, \href
  {https://ui.adsabs.harvard.edu/abs/2017ApJ...838...65P} {838, 65}

\bibitem[\protect\citeauthoryear{{Patra}, {Subrahmanyan}, {Sethi}, {Udaya
  Shankar}  \& {Raghunathan}}{{Patra} et~al.}{2015}]{2015ApJ...801..138P}
{Patra} N.,  {Subrahmanyan} R.,  {Sethi} S.,  {Udaya Shankar} N.,
  {Raghunathan} A.,  2015, \mn@doi [\apj] {10.1088/0004-637X/801/2/138}, \href
  {http://adsabs.harvard.edu/abs/2015ApJ...801..138P} {801, 138}

\bibitem[\protect\citeauthoryear{{Patwa}, {Sethi}  \& {Dwarakanath}}{{Patwa}
  et~al.}{2021}]{patwa21}
{Patwa} A.~K.,  {Sethi} S.,   {Dwarakanath} K.~S.,  2021, \mn@doi [\mnras]
  {10.1093/mnras/stab989}, \href
  {https://ui.adsabs.harvard.edu/abs/2021MNRAS.504.2062P} {504, 2062}

\bibitem[\protect\citeauthoryear{{Planck Collaboration} et~al.,}{{Planck
  Collaboration} et~al.}{2014}]{2014A&A...571A..17P}
{Planck Collaboration} et~al., 2014, \mn@doi [\aap]
  {10.1051/0004-6361/201321543}, \href
  {http://adsabs.harvard.edu/abs/2014A%26A...571A..17P} {571, A17}

\bibitem[\protect\citeauthoryear{{Price} et~al.,}{{Price}
  et~al.}{2018}]{leda18}
{Price} D.~C.,  et~al., 2018, \mn@doi [\mnras] {10.1093/mnras/sty1244}, \href
  {https://ui.adsabs.harvard.edu/abs/2018MNRAS.478.4193P} {478, 4193}

\bibitem[\protect\citeauthoryear{{Pritchard} \& {Loeb}}{{Pritchard} \&
  {Loeb}}{2012}]{pritchard12}
{Pritchard} J.~R.,  {Loeb} A.,  2012, Reports on Progress in Physics, \href
  {http://adsabs.harvard.edu/abs/2012RPPh...75h6901P} {75, 086901}

\bibitem[\protect\citeauthoryear{{Reis}, {Fialkov}  \& {Barkana}}{{Reis}
  et~al.}{2020}]{reis20}
{Reis} I.,  {Fialkov} A.,   {Barkana} R.,  2020, \mn@doi [\mnras]
  {10.1093/mnras/staa3091}, \href
  {https://ui.adsabs.harvard.edu/abs/2020MNRAS.499.5993R} {499, 5993}

\bibitem[\protect\citeauthoryear{{Reis}, {Fialkov}  \& {Barkana}}{{Reis}
  et~al.}{2021}]{reis21}
{Reis} I.,  {Fialkov} A.,   {Barkana} R.,  2021, \mn@doi [\mnras]
  {10.1093/mnras/stab2089}, \href
  {https://ui.adsabs.harvard.edu/abs/2021MNRAS.506.5479R} {506, 5479}

\bibitem[\protect\citeauthoryear{{Ross}, {Dixon}, {Ghara}, {Iliev}  \&
  {Mellema}}{{Ross} et~al.}{2019}]{2019MNRAS.487.1101R}
{Ross} H.~E.,  {Dixon} K.~L.,  {Ghara} R.,  {Iliev} I.~T.,   {Mellema} G.,
  2019, \mn@doi [\mnras] {10.1093/mnras/stz1220}, \href
  {https://ui.adsabs.harvard.edu/abs/2019MNRAS.487.1101R} {487, 1101}

\bibitem[\protect\citeauthoryear{{Ross}, {Giri}, {Mellema}, {Dixon}, {Ghara}
  \& {Iliev}}{{Ross} et~al.}{2021}]{2021MNRAS.506.3717R}
{Ross} H.~E.,  {Giri} S.~K.,  {Mellema} G.,  {Dixon} K.~L.,  {Ghara} R.,
  {Iliev} I.~T.,  2021, \mn@doi [\mnras] {10.1093/mnras/stab1822}, \href
  {https://ui.adsabs.harvard.edu/abs/2021MNRAS.506.3717R} {506, 3717}

\bibitem[\protect\citeauthoryear{{Rybicki} \& {Lightman}}{{Rybicki} \&
  {Lightman}}{1986}]{rybicki86}
{Rybicki} G.~B.,  {Lightman} A.~P.,  1986, {Radiative Processes in
  Astrophysics}.
John Wiley \& Sons, Ltd

\bibitem[\protect\citeauthoryear{{Seager}, {Sasselov}  \& {Scott}}{{Seager}
  et~al.}{2000}]{Seager2000}
{Seager} S.,  {Sasselov} D.~D.,   {Scott} D.,  2000, \mn@doi [\apjs]
  {10.1086/313388}, \href
  {https://ui.adsabs.harvard.edu/abs/2000ApJS..128..407S} {128, 407}

\bibitem[\protect\citeauthoryear{{Singh} et~al.,}{{Singh}
  et~al.}{2018}]{singh18}
{Singh} S.,  et~al., 2018, \mn@doi [\apj] {10.3847/1538-4357/aabae1}, \href
  {https://ui.adsabs.harvard.edu/abs/2018ApJ...858...54S} {858, 54}

\bibitem[\protect\citeauthoryear{{Sokolowski} et~al.,}{{Sokolowski}
  et~al.}{2015}]{2015PASA...32....4S}
{Sokolowski} M.,  et~al., 2015, \mn@doi [\pasa] {10.1017/pasa.2015.3}, \href
  {http://adsabs.harvard.edu/abs/2015PASA...32....4S} {32, e004}

\bibitem[\protect\citeauthoryear{{Thomas} et~al.,}{{Thomas}
  et~al.}{2009}]{Thom09}
{Thomas} R.~M.,  et~al., 2009, \mn@doi [\mnras]
  {10.1111/j.1365-2966.2008.14206.x}, \href
  {http://adsabs.harvard.edu/abs/2009MNRAS.393...32T} {393, 32}

\bibitem[\protect\citeauthoryear{{Trott} et~al.,}{{Trott}
  et~al.}{2020}]{trott20}
{Trott} C.~M.,  et~al., 2020, \mn@doi [\mnras] {10.1093/mnras/staa414}, \href
  {https://ui.adsabs.harvard.edu/abs/2020MNRAS.493.4711T} {493, 4711}

\bibitem[\protect\citeauthoryear{{Villanueva-Domingo}, {Mena}  \&
  {Miralda-Escud{\'e}}}{{Villanueva-Domingo} et~al.}{2020}]{pablo20}
{Villanueva-Domingo} P.,  {Mena} O.,   {Miralda-Escud{\'e}} J.,  2020, \mn@doi
  [\prd] {10.1103/PhysRevD.101.083502}, \href
  {https://ui.adsabs.harvard.edu/abs/2020PhRvD.101h3502V} {101, 083502}

\bibitem[\protect\citeauthoryear{{Voytek}, {Natarajan}, {J{\'a}uregui
  Garc{\'{\i}}a}, {Peterson}  \& {L{\'o}pez-Cruz}}{{Voytek}
  et~al.}{2014}]{2014ApJ...782L...9V}
{Voytek} T.~C.,  {Natarajan} A.,  {J{\'a}uregui Garc{\'{\i}}a} J.~M.,
  {Peterson} J.~B.,   {L{\'o}pez-Cruz} O.,  2014, \mn@doi [\apjl]
  {10.1088/2041-8205/782/1/L9}, \href
  {http://adsabs.harvard.edu/abs/2014ApJ...782L...9V} {782, L9}

\bibitem[\protect\citeauthoryear{{Xu}, {Yue}  \& {Chen}}{{Xu}
  et~al.}{2021}]{xu21}
{Xu} Y.,  {Yue} B.,   {Chen} X.,  2021, arXiv e-prints, \href
  {https://ui.adsabs.harvard.edu/abs/2021arXiv210212865X} {p. arXiv:2102.12865}

\bibitem[\protect\citeauthoryear{{Zaroubi}}{{Zaroubi}}{2013}]{2013ASSL..396...45Z}
{Zaroubi} S.,  2013, in {Wiklind} T.,  {Mobasher} B.,   {Bromm} V.,  eds,
  Astrophysics and Space Science Library Vol. 396, The First Galaxies. p.~45
  (\mn@eprint {arXiv} {1206.0267}), \mn@doi{10.1007/978-3-642-32362-1_2}

\bibitem[\protect\citeauthoryear{de Lera~Acedo}{de~Lera~Acedo}{2019}]{8879199}
de Lera~Acedo E.,  2019, in 2019 International Conference on Electromagnetics
  in Advanced Applications (ICEAA). pp 0626--0629,
  \mn@doi{10.1109/ICEAA.2019.8879199}

\makeatother
\end{thebibliography}






\bsp	
\label{lastpage}
\end{document}